\documentclass{aa}
\usepackage{graphicx}

\def\arcmin{\hbox{$^\prime$}}
\def\arcsec{\hbox{$^{\prime\prime}$}}

\def\farcm{\hbox{$.\mkern-4mu^\prime$}}
\def\farcs{\hbox{$.\!\!^{\prime\prime}$}}

\begin{document}
%

%
\title{Tip of the Red Giant Branch Distance for the Sculptor Group Dwarf 
ESO~540-032
\thanks{Based on observations collected at the European Southern Observatory 
(ESO 64.N-0069)}
}

\author{Helmut Jerjen\,\inst{1}
\and Marina Rejkuba\,\inst{2,3}} 

\offprints{H.~Jerjen, e-mail: jerjen@mso.anu.edu.au}   

\institute{Research School of Astronomy and Astrophysics, 
The Australian National University, Mt Stromlo Observatory, Cotter Road, 
Weston ACT 2611, Australia,
jerjen@mso.anu.edu.au
\and 
European Southern Observatory,  Karl-Schwarzschild-Strasse 2, 
D-85748 Garching bei M\"unchen, Germany,
mrejkuba@eso.org
\and 
Department of Astronomy, P. Universidad Cat\'olica, Casilla 306, 
Santiago 22, Chile
}

\date{Received 19 January 2001; accepted 9 March 2001}
\authorrunning{Jerjen \& Rejkuba}
\titlerunning{TRGB Distance for ESO 540-032}  

\abstract{
We present the first $VI$ CCD photometry for the Sculptor group galaxy ESO 540-032
obtained at the Very Large Telescope UT1+FORS1. The $(I,\, V-I)$ colour-magnitude 
diagram indicates that this intermediate-type dwarf galaxy is dominated by old, 
metal-poor ([Fe/H]$\approx -1.7$\,dex) stars, with a small population of slightly 
more metal-rich ([Fe/H]$\approx -1.3$\,dex), young (age $150-500$\,Myr) stars. A 
discontinuity in the $I$-band luminosity function is detected at $I_0$ = $23.44\pm$0.09\,mag. 
Interpreting this feature as the tip of the red giant branch and adopting 
$M_{I} = -4.20\pm 0.10$\,mag for its absolute magnitude, we have determined a Population 
II distance modulus of $(m - M)_{0} = 27.64 \pm 0.14$\,mag $(3.4 \pm 0.2$\,Mpc). 
This distance confirms ESO 540-032 as a member of the nearby Sculptor group but is 
significantly larger than a previously reported value based on the Surface Brightness 
Fluctuation (SBF) method. The results from stellar population synthesis models suggest 
that the application of the SBF technique on dwarf galaxies with mixed morphology 
requires a detailed knowledge of the underlying stellar composition and thus offers no 
advantage over a direct distance measurement using the tip of the red giant branch as 
distance indicator. We produce the surface brightness profiles for ESO 540-032 and 
derive the photometric and structural parameters. The global properties follow closely 
the relations between metallicity and both absolute magnitude and central surface 
brightness defined by dwarf elliptical galaxies in the Local Group. Finally, we identify 
and discuss a non-stellar object near the galaxy center which may resemble a globular 
cluster.
}

\maketitle

\section{Introduction}
Dwarf elliptical galaxies (hereafter dEs, subsuming ``dwarf spheroidal galaxies'') in 
the Local Group exhibit a rather diverse and complex set of star formation histories 
(Da Costa 1998; Grebel 1998). Their stellar populations range from almost 
exclusively old (Ursa Minor, Olszewski \& Aaronson 1985) to mainly old 
(e.g.~Tucana, Da Costa 1998) with an observed intermediate-age episodic SFH  
in Carina (Mighell 1997; Hurley-Keller et al.~1998) and an intermediate-age 
continuous SFH in Fornax (Stetson et al.~1998). Phoenix and LSG3 are 
classified as intermediate-type dwarfs (dE/Im), because they show similarities to 
both dwarf ellipticals and dwarf irregulars. These systems are dominated by an old 
metal-poor population with no evidence for major star formation activities after the 
initial episode 8--10\,Gyr ago. However, both systems also accommodate a minor population 
of young stars, with ages of about 150--500\,Myr, which makes these galaxies resemble 
dwarf irregulars (Held et al.~1999; Mart\'{\i}nez-Delgado et al.~\cite{delgado99}; 
Holtzman et al.~2000; Aparicio et al.~1997). 

This rich collection of star formation histories in the Local Group led in recent 
years to a large effort to find and physically characterise more of these 
faintest, most elusive galaxies beyond the group boundary. An extensive search 
for dwarf galaxies in the closest galaxy aggregate to the Local Group, the 
Sculptor (Scl) Group at $\approx 2.5$\,Mpc, was carried out by C\^ot\'e 
and collaborators (1995; C\^ot\'e et al.~1997) confirming 16 gas-rich dwarf 
members based on pointed H\,I and H$\alpha$ observations and redshift measurements. 
Her list was supplemented by a first set of five gas-poor dE and intermediate-type 
dwarf galaxies (Jerjen et al.~1998, hereafter JFB98; 2000b) whose distances were 
estimated with the Surface Brightness Fluctuation (SBF) method. Thereby, the zero 
point for the reported SBF distances had to rely on theoretical predictions
of stellar population synthesis models and on assumptions about the dwarfs' 
stellar contents and star formation histories. The lack of further distance 
information initiated our work to get a distance confirmation for some of 
these dwarf galaxies based on their resolved stellar populations. 

\begin{figure*}
\centering
\includegraphics[width=8cm,angle=270]{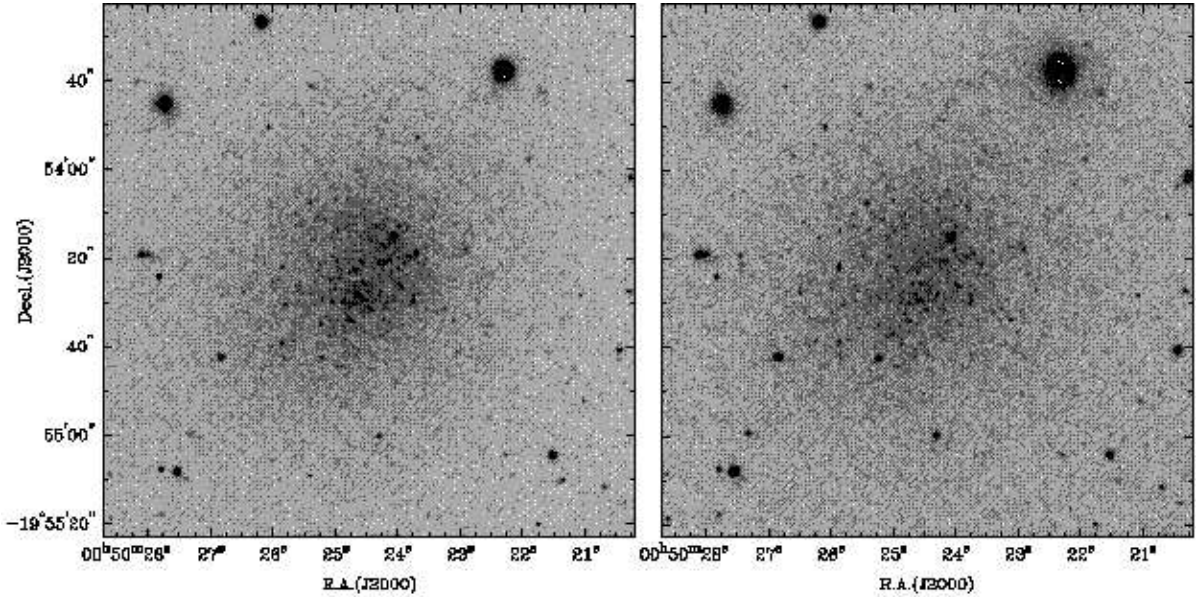}
\caption[F1.ps]{
$V$ (left) and $I$-band (right) images of ESO 540-032, taken at the VLT (UT1+FORS1) 
and showing resolution into stars. The Field is $3\farcm5 \times 3\farcm5$. 
North is at the top and east to the left. \label{galimas}}
\end{figure*}

Among these new Scl group dwarfs was ESO 540-032, also known as KK010 
(Karachentseva \& Karachentsev 1998). This galaxy was initially classified as 
intermediate-type dwarf, dE/Im (JFB98) or equivalent Sph/dIrr (Karachentseva \& 
Karachentsev 1998) based on its dE-like overall smooth light distribution and the 
knotty structure, typical for members of the dwarf irregular family. These 
morphological features are visible in Fig.~\ref{galimas} showing $V$ and $I$ CCD 
images of the galaxy. ESO 540-032 remained undetected in the pointed \ion{H}{i} 
surveys by Gallagher et al.~(\cite{gallagher95}), C\^ot\'e et al.~(1997) and 
Karachentsev et al.~(\cite{kara99}). C\^ot\'e et al. quote a $3\sigma$ detection 
limit of $4\times 10^6$\,M$_\odot$\,\ion{H}{i} mass which is a factor of 10 above 
the expected \ion{H}{i} mass for other such dwarf systems (St-Germain et al.~1999). 

In this paper we present the first $VI$ CCD photometry of the stellar content 
of the Scl group dwarf ESO 540-032. In \S2 we describe the observations, data 
reduction, and the photometry completeness tests. In \S3 we analyse the 
resulting colour-magnitude diagram. The tip magnitude of the red giant 
branch is measured and the distance of the galaxy determined based on this 
method in \S4. The latter result is compared with the SBF distance in \S5. Finally, 
surface brightness profiles and parameters from S\'ersic model fitting are derived 
in \S6 and a globular cluster candidate discussed in \S7. 

\section{Observations and Reductions}
\label{data}
\begin{table*}
\caption[]{Summary of Observations}
\label{obslog}
\begin{tabular}{cccccccc}
\hline \hline
R.A. (J2000) & Decl.~(J2000) & Date & Telescope \&  & Exposure & Filter & Airmass & Seeing \\
(h min sec)       & ($^\circ$  $\arcmin$  $\arcsec$)&  (dd/mm/yy)
& Instrument &  (sec)   &        &        & ($\arcsec$)\\
\hline \hline
00 50 18 &$-$19 55 45&13/11/2000&Antu+FORS1  &3$\times$660& $V$ & 1.25 &0.56\\
00 50 18 &$-$19 55 45&07/01/2000&Antu+FORS1  &3$\times$550& $I$ & 1.20 &0.67\\
\hline
\end{tabular}
\end{table*}

Deep CCD images of ESO 540-032 were acquired in service mode using FORS1 (FOcal 
Reducer and low dispersion Spectrograph) at Antu (UT1) VLT at ESO Paranal 
Observatory. The FORS1 detector is a 2048$\times$2048 Tektronix CCD, thinned 
and anti-reflection coated. The pixel size is 24$\times$24 $\mu$m.
The field of view was $6\farcm8 \times 6\farcm8$ and the pixel scale 
$0\farcs2/$pixel. 

Three 660\,sec exposures were taken in the $V$-band during the night of 
1999 November 13, and three 550\,sec $I$-band exposures were secured during the 
night of 2000 January 7. The summary of observations specifying air masses and 
seeing conditions is given in Table \ref{obslog}. Photometric standard stars from 
the lists of Landolt (\cite{landolt}) and Walker (\cite{walker}) were observed 
in both $V$ and $I$ filters, during the first and the second night, respectively.

In service mode, direct images are read out by default in the four-port mode, i.e. 
four amplifiers read out one quadrant of the CCD each. Xccdred, an 
IRAF\footnote{IRAF is distributed by the National 
Optical Astronomy Observatories, which is operated by the Association of Universities 
for Research in Astronomy, Inc., under contract with the National Science Foundation} 
package specially developed to reduce such images, was used to subtract overscan for 
each amplifier area individually and later to subtract the bias and flat-field the images. 
The three images in each filter were then registered using {\it imalign} and averaged 
with {\it imcombine}. Finally, the {\it crreject} algorithm was employed to 
reject cosmic rays. 

\subsection{Photometric Calibrations}
In service mode at Paranal, a set of photometric standard stars from the Landolt 
(\cite{landolt}) catalogue are observed in various passbands each night. On November 13, 
two Landolt fields, totalling 12 stars were observed and on January 7, four Landolt 
fields, totalling 19 stars. In order to have more standards that cover a broader range 
in colour and in
brightness, we further observed a standard star sequence close to NGC~300 that was 
established by Graham (\cite{graham81}) and Walker (\cite{walker}). In this way a total 
of 28 and 35 standards could be used to derive extinction coefficients and zero points of 
the two nights separately. The derived photometric transformations show that the two nights 
had different zero points. Since our scientific observations in $V$ and $I$-band were
taken on different nights, we calibrated the data in two steps.

First, the zero point and extinction coefficients were calculated for each
night separately using the equation:
\begin{equation}
m_{\rm inst} = {\rm ZP} + k \times {\rm Airmass} + m_{\rm cat}
\label{zp} 
\end{equation}
where $m_{\rm inst}$ and $m_{\rm cat}$ are the instrumental and catalogued
magnitudes, ZP is the zero point and $k$ is the extinction coefficient. 
We defined magnitudes that were calibrated with this zero order correction for
atmospheric effects as $m_{\rm fit}=m_{\rm inst} - {\rm ZP}- k \times {\rm Airmass}$.
 
After we determined the zero point and extinction coefficient, the
dependence of the calibration on the colour term was explored. It turned out 
that the colour term was important for the $V$-band calibration but negligible 
for the $I$-band. This allowed to put $I_{\rm real} = I_{\rm fit}$ (see also 
the value of colour coefficient $c$ in Table~\ref{calibration_coeff}). 

The second step in calibration involved the equation of the form for $V$-band:

\begin{equation}
V_{\rm real} - V_{\rm fit} = c \times (V-I)_{\rm fit} + b
\end{equation}

where $V_{\rm real}$ is the magnitude from the standard star catalogue, 
$V_{\rm fit}$ is the magnitude computed after the zero point and extinction 
corrections (Eq.\ref{zp}), $c$ is the colour coefficient and $b$ is the 
constant of the linear fit. The colour term depends on telescope and detector 
characteristics and should be constant in time.
We computed the colour term for $V$-band for both nights separately and with
the combined data and the three values are almost identical and well within
the errors of the fit. 
The values of ZP, $k$, $c$ and $b$ for $V$-band and $I$-band calibration 
are reported in Table~\ref{calibration_coeff}. We note that the given $c$ 
and $b$ values are mean values for the two nights.

\begin{table*}
\caption[]{Photometric Calibration Coefficients}
\label{calibration}
\begin{tabular}{cccccccccc}
\hline \hline
Night & Filter & ZP & $\sigma_{\rm ZP}$ & $k$ & $\sigma_{k}$ & $c$ & $\sigma_{c}$ & 
$b$ & $\sigma_{b}$ \\
\hline
\hline
1999 Nov 13 & $V$ & $-$27.276 & 0.07 & 0.105 & 0.06& 0.077 & 0.013 & $-$0.041 & 0.013 \\ 
2000 Jan 07 & $I$ & $-$26.292 & 0.07 & 0.138 & 0.06 & $-$0.007 & 0.015& 0.009 & 0.015\\
\hline
\end{tabular}
\label{calibration_coeff}
\end{table*}

\begin{figure}
\centering
\resizebox{\hsize}{!}{\includegraphics{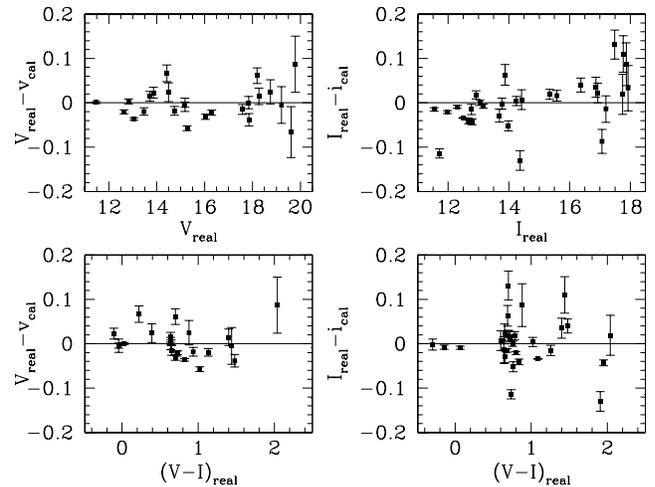}}
\caption[F2.ps]{Difference between the real and observed (calibrated) magnitudes of
standard stars as a function of magnitude (upper two panels) and colour
(bottom panels). Error bars represent combined errors of photometry and
errors of the catalogued magnitudes.}
\label{calib_fig}
\end{figure}

Applying the transformations described above reproduce the standard
magnitudes with a rms of 0.04 in $V$ and 0.05 in $I$. Fig.~\ref{calib_fig}
shows the difference between the real and calibrated magnitudes of
standard stars as a function of magnitude and colour.

\begin{figure}
\centering
\resizebox{\hsize}{!}{\includegraphics{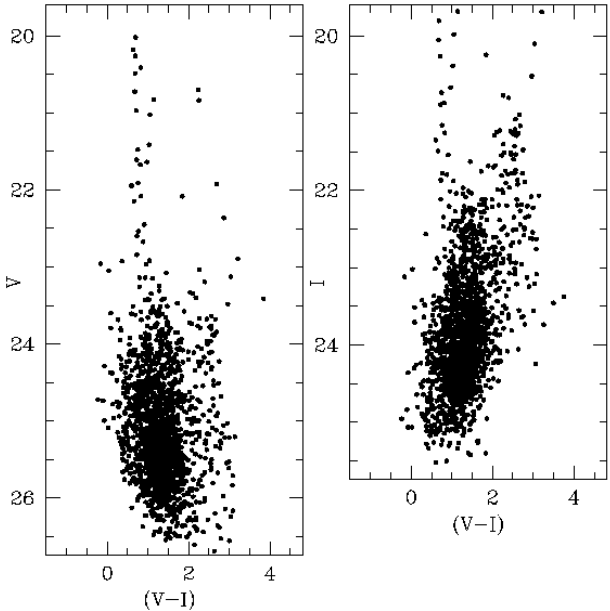}}
\caption[F3.ps]{Colour-magnitude diagrams of all 1830 stars 
matched in the $V$ and $I$-band CCD images.}
\label{CMDall}
\end{figure}

The photometric measurements were performed within the IRAF environment
using DAOPHOT~II (Stetson~\cite{stetson}). All the objects in the field with
more than 3$\sigma$ above the background level were located separately in 
the $V$ and $I$ frames. A variable point-spread function (PSF) was 
constructed by using 32 and 33 stars in the $V$ and $I$ frames, respectively.  
The magnitudes were then measured using the {\it allstar} task. Only objects 
that matched in both $V$ and $I$ band with good photometry (i.e. $\sigma_{\rm mag} 
< 0.5$, $-1< sharp <1$ and $ chi < 2$) were retained. Restricting the $sharp$ 
and $chi$ parameters excludes extended objects like galaxies and stellar 
blends as well as any remaining blemishes. The resulting $(V,V-I)$ and $(I,V-I)$ 
colour-magnitude diagrams consisting of all 1830 stars detected on the CCD are 
shown in Fig.~\ref{CMDall}. 

\subsection{Completeness and Contamination}
\begin{figure}
\centering
\resizebox{\hsize}{!}{\includegraphics{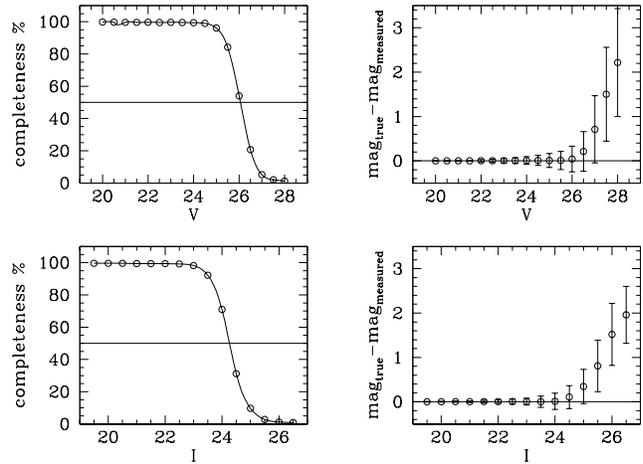}}
\caption[F4.ps]{
Artificial-star tests: completeness as function of magnitude 
(left panels) and photometric accuracy as a function of magnitude 
(right panels) for the $V$ and $I$ band.}
\label{complet_mag}
\end{figure}

We made extensive tests to measure completeness and magnitude errors as
function of magnitude and radial distance from the center of the galaxy.
Using the {\it addstar} routine in DAOPHOT~II (Stetson~\cite{stetson}) we added
400 stars at a time to $V$ and $I$ frame, all with the same magnitude and
uniformly distributed over the whole field, and re-computed the photometry.
This operation was repeated for 5 different positions of added stars and at
each 0.5 magnitudes in order to have a statistically significant sample of
2000 added stars per magnitude bin.
Incompleteness, defined as a recovery rate of 50\% for the stars from the
artificial-star experiment, sets in at $V=26.1$ and $I=24.3$. The latter value 
is well below the magnitude of the tip of the red giant branch, i.e. 
$I\approx$23.5 (see Sect.~4) at which level our photometry is complete 
more than 90\% (Fig.~\ref{complet_mag}).

\begin{figure}
\centering
\resizebox{\hsize}{!}{\includegraphics{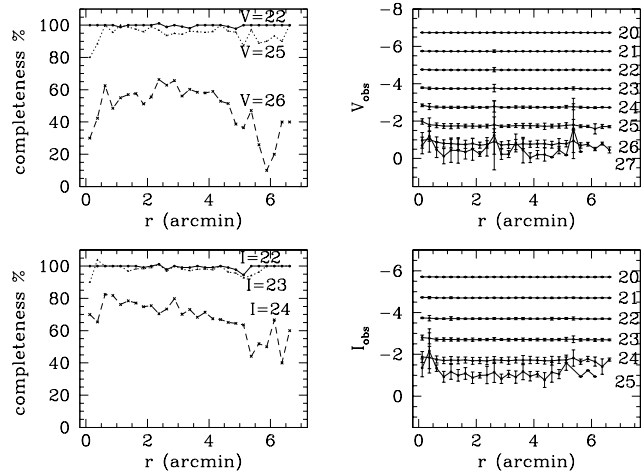}}
\caption[F5.ps]{
Artificial-star tests: completeness and photometric accuracy as
function of radial distance from the center of ESO 540-032. On the right
panel Y-axis is the instrumental observed magnitude, while the label
indicates the corresponding calibrated magnitudes.}
\label{complet_rad}
\end{figure}

The artificial-star tests were performed also to assess the accuracy of 
our photometry. The difference between the recovered and input magnitude 
as a function of magnitude and as a function of radial distance from the 
galaxy center are shown on the right panels of Figs.~\ref{complet_mag} 
and ~\ref{complet_rad}. From these experiments we conclude that our 
photometry is reliable down to the incompleteness limit. The measured 
magnitudes of fainter stars are systematically brighter. 

On the left panels of Fig.~\ref{complet_rad} the completeness as function
of radial distance is shown for three different input magnitudes, at
$\sim$100, 90 and 50\% completeness levels. For brighter magnitudes there is
no radial dependence on completeness, nor systematic differences in input
and recovered magnitudes (right panels), but for magnitudes close to 90\%
completeness level the inner most bin, the circle within the 0.125\,arcmin
from the galaxy center, has significantly lower completeness and the
recovered magnitudes start to be systematically brighter because of blending
effects due to crowding. Blending with other stars results in
brightening of the measured magnitudes by $\sim$0.1\,mag at the level of 90\%
completeness and only for the inner most radial bin. 

There is only little foreground contamination from the Galactic halo and disk 
due to low galactic latitude of the galaxy (l$=121.01^\circ$, b$=-82.77^\circ$).
Most of the foreground stars appear as a plume of bright blue stars
around $V-I\sim0.7$ (see Fig.~\ref{CMDall}). 
We used the Besanc\'on group model of stellar population synthesis of the Galaxy
(Robin et al.~\cite{robin}) to estimate the total number of galactic
foreground stars in our field. They amount to 84 stars in our observed
magnitude range ($20<V<26.5$), and $~1/3$ of them are mixed with
ESO 540-032 stars. However, we are interested in the restricted 
magnitude range around the red giant branch (RGB) tip where only few 
foreground stars could be found.

Most of the background galaxies are extended and rejected by the $sharp$ and $chi$ 
parameter constrains in our photometry, but some compact galaxies might 
still contaminate the sample. Since the optical size of ESO 540-032 is much smaller 
than the FORS1 field of view, we centered the galaxy on the upper left quadrant. 
In that way we can also estimate the foreground stellar and background galaxy 
contamination by counting the number of objects in our final photometry list 
that are detected in the lower right corner of the CCD. 
There are 121 objects in the area corresponding to $\approx 1/4$ of CCD. Some of 
them belong to the halo of ESO 540-032, but mostly they are galactic stars and 
compact background galaxies. 

\begin{figure*}
\centering
\includegraphics[width=17cm]{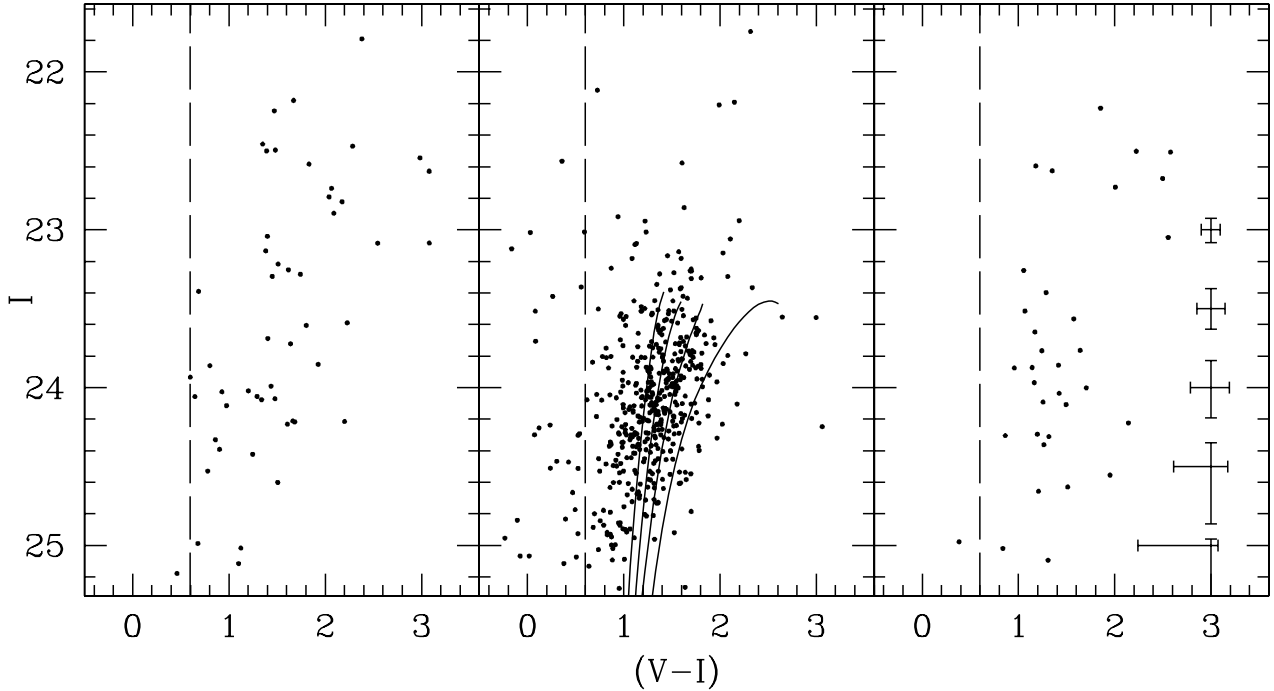}
\caption[F6.ps]{
The central panel shows the $I$-band colour-magnitude diagram of all stars matched 
in $V$ and $I$-band images and within the 40 arcsec radius from the galaxy center. 
Solid lines from blue to red represent the standard globular cluster branches 
for M15 ([Fe/H]=$-2.17$), M2 ([Fe/H]=$-1.58$), NGC 1851 ([Fe/H]=$-1.16$), and 
47\,Tuc ([Fe/H]=$-0.71$) taken from Da Costa \& Armandroff (1990) and 
shifted to the distance modulus and reddening of ESO 540-032. For 
comparison we show in the left and right panel the CMDs for two comparison fields 
produced from an equivalent area on the CCD. To illustrate the presence of a significant
blue stellar population in the galaxy, a dashed line is drawn at 
(V-I)=0.6 in each panel. The error bars reflect the photometric uncertainties and
systematic errors.}   
\label{CMDgal}
\end{figure*}

\section{Colour-magnitude diagram and stellar contents}
The central panel of Fig.~\ref{CMDgal} shows the $(I,V-I)$ colour-magnitude diagram (CMD) 
produced from all 469 stars identified within a 40 arcsec radius from 
the galaxy center, an area which covers the main body of ESO 540-032. The most prominent 
feature is a well-defined red giant branch with an approximate tip magnitude 
$I\approx 23.5$. The colour dispersion of the RGB stars suggests a wide range in metallicity. 
Superimposed on the CMD are RGB fiducials from Da Costa \& Armandroff (\cite{DA90}) 
for the Galactic globular clusters M15 ([Fe/H]$=-2.17$), M2 ([Fe/H]$=-1.58$), NGC 1851 
([Fe/H]$=-1.16$), and 47\,Tuc ([Fe/H]$=-0.71$). The giant branches are reddened by E$(B-V)=0.020$\,mag 
(Schlegel et al.~1998) and brought to the distance of ESO 540-032 using the distance 
modulus of 27.68\,mag (see Sect.~4) in order to match the observed CMD. The majority 
of stars are metal-poor with $-2.17<$ [Fe/H] $<-1.16$. There are also a few stars as 
metal-rich as 47\,Tuc but almost none are more metal-rich. Caldwell et al.~(1998)
presented a quadratic relation between the mean metallicity and the dereddened giant 
branch colour $(V-I)_{0,-3.5}$ at $M_I=-3.5$: $\langle$[Fe/H]$\rangle=-1.00 + 
1.97q -3.2 q^2$, where $q=[(V-I)_{0,-3.5}-1.6]$. Using this equation with the median 
colour $(V-I)_{0,-3.5}=1.34\pm 0.04$ for stars in the magnitude interval 
$24.13<I<24.23$ (corresponding to $M_I=-3.5\pm0.05$ for ESO 540-032), we derive 
$\langle$[Fe/H]$\rangle=-1.7(\pm0.3)$\,dex where the given error includes uncertainties 
in the distance modulus, photometry, reddening, and a rms of 0.08\,dex from the 
relation.

A noticeable number of stars are brighter than the tip magnitude. They could be 
intermediate-age AGB stars. However, CMDs of two comparison fields on the CCD 
with the same area as the galaxy field and situated $\approx 4$\,armin west and south of 
the galaxy center (left and right panel of Fig.~\ref{CMDgal}) show a similar amount 
of foreground stars in the critical magnitude range $22 < I < 23.5$. This field 
contamination prevents a detailed discussion of the existence or a number estimate 
of AGB stars in ESO 540-032 with the current photometric data.   

The nature of ESO 540-032 being an intermediate-type dwarf galaxy is
revealed by the presence of a small population of blue stars. The panels 
in Fig.~\ref{CMDgal} illustrate that stars bluer than $(V-I)\approx 0.6$ 
are found most exclusively in the galaxy area (only one such star appears in 
each comparison field) thus a confusion by foreground stars can be ruled out. 
These stars are likely to be the bluest extent of the He burning phase for 
metal-poor stars with $-1.7\leq$[Fe/H]$\leq-1.3$ and ages in the range $150-500$\,Myr
(see Fig.~13 of Mart\'{\i}nez-Delgado et al.~\cite{delgado99}). 
Overall, the CMD of ESO 540-032, as far as it has been recovered 
($\approx 1.5$\,mag below the RGB tip), has features in common with 
the CMDs of the intermediate-type dwarfs Phoenix (Held et al.~1999;
Mart\'{\i}nez-Delgado et al.~\cite{delgado99}; Holtzman et al.~2000) 
and LSG3 (Aparicio et al.~1997). 

\begin{figure}
\centering
\resizebox{\hsize}{!}{\includegraphics{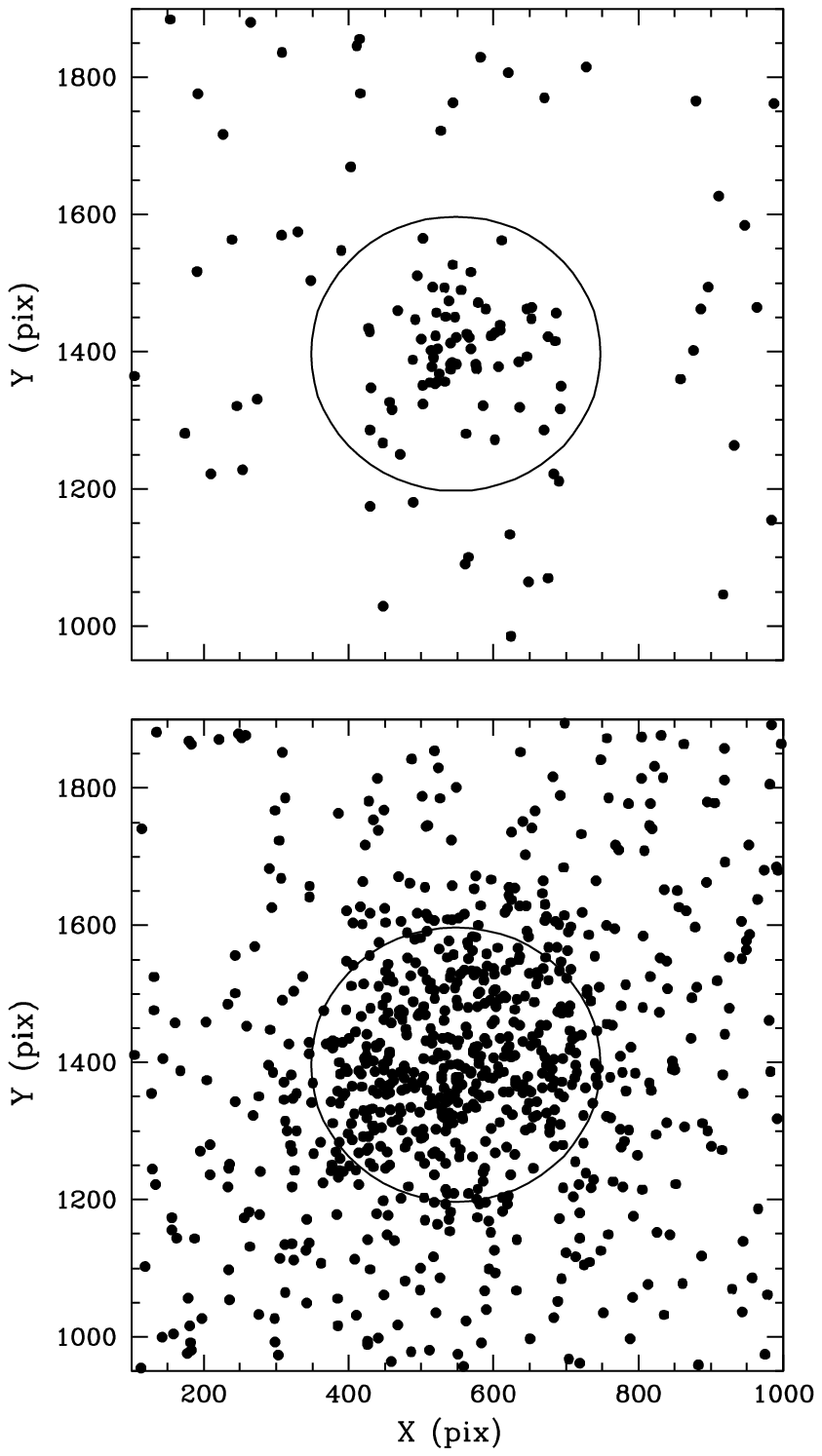}}
\caption[F7.ps]{
Pixel maps of all detected stars in the galaxy and its vicinity. The 
circles are indicating the 40 arcsec boundary ($\mu_I\approx 25.0$\,mag\,arcsec$^{-2}$) 
of the galaxy. Stars bluer than $(V-I)=0.9$ (top panel) show a higher
concentration to the center than the red stars (bottom panel). 
\label{galstardist}}
\end{figure}

Finally, we look at the distribution of the stars in the galaxy
(Fig.~\ref{galstardist}). All stars bluer than $(V-I) = 0.9$ appear 
strongly concentrated in the galaxy center with a rapid decline 
in numbers to larger radii. A less steep gradient is found for the 
red stars which is consistent with the results from the 
surface brightness profiles of ESO 540-032 (see Sect.~5 and Jerjen et 
al.~2000b) where the half light radius is measured to be significantly 
smaller in $B$ than in $I$ ($r_{B, \rm eff}=22.6\pm0.9$\,arcsec
versus $r_{I, \rm eff}=29.4$\,arcsec). The difference in the 
stellar distributions suggest the most recent star formation
activity took place close to the center of the galaxy.

\section{Tip of the Red Giant Branch Distance}
The tip magnitude of the red giant branch (TRGB) which marks the core
helium flash of low-mass Population II stars has been proven 
to be a reliable and accurate distance indicator. Original work on the 
empirical calibration of the method was based on observations of 
giant branches for standard Galactic globular clusters (Da Costa \& 
Armandroff \cite{DA90}). A more recent study by Gratton et al.~(\cite{gratton}) 
used the Hipparcos subdwarf parallax distance scale to recalibrate
these data. The absolute magnitude of the tip in the $I$-band 
was found to be virtually constant at $M_I=-4.2$\,mag for ages 
2--15\,Gyr and for metallicities between $-2.2<$[Fe/H]$<-0.7$
with a scatter of only $\pm 0.1$\,mag. These empirical results are 
well supported by the theoretical RGB models of Salaris \& Cassisi (1998).

\begin{figure}
\centering
\resizebox{\hsize}{!}{\includegraphics{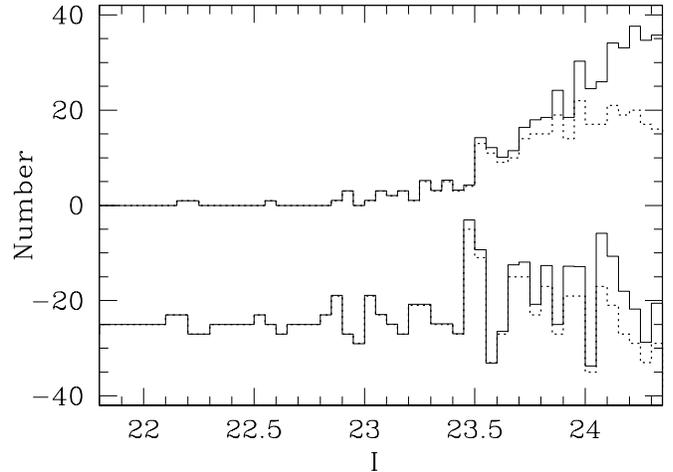}}
\caption[F8.ps]{
The top histograms show the observed (dashed line) and completeness-corrected
(solid line) $I$-band luminosity functions for the red giant stars $(V-I>0.9)$ 
in ESO 540-032. The convolutions of the luminosity functions with the Sobel 
edge-detection filter are the two corresponding bottom lines. The maxima of these 
edge-detector functions occur at the same magnitude $I\approx 23.5$\,mag and 
mark the steepest rise in the luminosity functions, i.e. the tip of the red giant
branch. The bin size is 0.05\,mag. 
\label{lf}}
\end{figure}

Since the TRGB method works for old and metal-poor stellar 
populations, we can use the well-defined RGB branch in the CMD of 
ESO 540-032 to measure the galaxy distance. The standard strategy 
is to determine the apparent magnitude of the tip which appears 
as a prominent local feature in the $I$-band luminosity function. 
For that purpose, the observed and completerness-corrected luminosity 
functions of all stars with $V-I>0.9$ (Fig.~\ref{lf}) are convolved 
with a zero-sum Sobel kernel [$-2, 0, 2$]. The resulting two edge-detection 
functions (Fig.~\ref{lf}) reach their maximum at the same magnitude, 
where the count discontinuity is the greatest. This technique of locating 
the tip in an objective manner was tested by Lee et al.~(\cite{lee93}). 
It provides consistent results independently of the chosen bin 
size or photometric precision. 

As the TRGB magnitude we use the midpoint of the corresponding magnitude 
bin $I_{\rm TRGB}=23.48\pm0.09$\,mag, where the error estimate is the 
combination of the width of the spike centered on the apparent magnitude 
of the TRGB ($\pm 0.05$\,mag) and the zero-point uncertainty of the 
photometry ($\pm 0.07$\,mag). Adopting $M_{I} = -4.2\pm 0.1$\,mag 
as the absolute magnitude and the reddening value $A_I=0.040\pm0.006$ 
from Schlegel et al.~(1998), we obtain for ESO 540-032 a distance modulus 
of $(m - M)_{0, {\rm TRGB}} = 27.64 \pm 0.14$\,mag $(3.4 \pm 0.2$\,Mpc). 

In principle, the start of the steep rise in the luminosity function 
could also be the consequence of a numerous AGB population. The RGB tip would 
be roughly one magnitude fainter at $I\approx 24.48$ where a clear detection is 
prevented by incompleteness. To test this possibility we assumed that the 
stars in the magnitude range $23.48<I<24.48$ are AGB stars. We then adopted 
a number-scaled version of the $I$-band luminosity function of Phoenix 
(Mart\'{\i}nez-Delgado et al.~\cite{delgado99}) to estimate the total light 
from all stars in ESO540-032. The resulting magnitude is more than 0.5\,mag 
brighter than the true value as derived in Sect.~6 based on aperture photometry.
Therefore the scenario of a large AGB population can be ruled out.

The distance of 3.4\,Mpc confirms ESO 540-032 as a member of the nearby, sparsely populated 
and complicate aggregate of galaxies we know as the Sculptor group. The elongated 
galaxy distribution of this group has a substantial depth, with member galaxies 
like NGC 55, ESO 294-010, and ESO 410-G005 at the near side at $\approx 1.7$\,Mpc, 
and NGC 45 and NGC 59 at the far side at $\approx 4.4$\,Mpc (JFB98; Karachentsev 
et al.~2000). In the sky, ESO 540-032 is close to the Sc galaxy NGC 247. However, 
their line-of-sight distances differ by 0.9\,Mpc which puts ESO 540-032 right 
in the field behind the spiral where the two dwarf irregulars DDO 006 (UGCA 015) 
and DDO 226 (IC 1574) were found (C\^ot\'e et al.~1997). 

\section{Comparison with Surface Brightness Fluctuation Distance}
The derived TRGB distance modulus for ESO 540-032 is significantly larger 
than the previous estimate based on the SBF method  at $(m-M)_{0,{\rm SBF}}=26.72 
\pm 0.13$\,mag or $2.2 \pm 0.14$\,Mpc (JFB98). In the latter study, the apparent 
fluctuation $R$-band magnitude $\bar{m_R}$ was recovered from the power spectrum 
of the galaxy image and converted into a distance modulus using predictions for 
$\bar{M_R}$ based on Worthey's (1994) stellar population synthesis models combined 
with Yale isochrones. Thereby, the underlying stellar population was assumed to be 
mainly old (8.5\,Gyr$<$ age $<$13\,Gyr) and metal-poor ($-1.9<$[Fe/H]$<-1.2$)
with a small pollution, up to the 10\% level in mass, of young (5\,Gyr), 
solar-metallicity stars. The modelling provided consistent results for such 
mixtures with $\bar{M_R}=-1.13\pm 0.06$\,mag. However, the one question 
remained about the influence of a young ($<8$\,Gyr) but {\em metal-poor} 
([Fe/H]$< -0.225$) polluting stellar population on the SBF zero point. 
As JFB98 pointed out, that part of the parameter space was not accepted 
by the model program. 

In the meantime, Worthey incorporated the isochrones from the Padova library 
(Bertelli et al.~1994) in his 
program\footnote{http://199.120.161.183:80/$\sim$worthey/dial/dial\_a\_model.html}
which not only yields significantly better predictions for $\bar{M_R}$ 
(for a discussion see Jerjen et al.~2000a) but also allows to test the still 
unexplored parameter space. Indeed, the results are quite different there. If 
for instance an old, metal-poor population (12\,Gyr, [Fe/H]$=-1.7$) is mixed 
(2\% in mass) with a young, slightly more metal-rich population (500\,Myr, 
[Fe/H]$=-1.3$), Worthey's models give $\bar{M_R}=-2.37$ with a colour of 
$(B-R)_0=0.98$. Using the newly derived zero point to recalibrate the fluctuation 
signals measured for the two fields F1 and F2 which have been analysed in 
JFB98 and having estimated colours of $(B-R)_{\rm 0,F1}=0.89$ and 
$(B-R)_{\rm 0,F2}=0.99$, we compute a distance modulus of $27.93\pm0.15$. This 
value now is in better agreement with the result from the TRGB analysis. 

Despite the apparent resolution of the distance discrepancy in the case of
ESO 540-032, we are not proposing to extend the application of the SBF technique 
to intermediate-type dwarfs. The value of $\bar{M_R}$ turns out to be highly variable
for such systems. It depends sensitively on the age and metallicity of the young 
stars as well as on the mass ratio $\mathcal{M}_{\rm young}/\mathcal{M}_{\rm old}$ 
of the two subpopulations.
The SBF is best measured in dwarf ellipticals and dwarf S0s, i.e. in dwarfs that show essentially 
no star formation subsequent to an initial formation episode. It seems very 
difficult if not impossible to predict the exact SBF zero point for an 
intermediate-type dwarf without profound knowledge of its stellar content. 
Thus, to obtain an accurate SBF distance would imply the analysis of a CMD which would 
naturally lead to a distance via a TRGB measurement.

The found discrepancy between the TRGB and SBF results for ESO 540-032 suggests 
the SBF distances for the two other intermediate-type Scl group dwarf galaxies ESO 294-010 
and ESO 540-030 are too small as well. While the location of ESO 294-010 at the 
near side of the Scl group seems to be confirmed by the low redshift at 
$v_\odot=117\pm 5$\,km\,s$^{-1}$ (JFB98), additional work will be needed to get 
a final answer in the case of ESO 540-030. 

\section{Galaxy Light Profiles and Model Parameters}
\begin{figure}
\centering
\resizebox{\hsize}{!}{\includegraphics{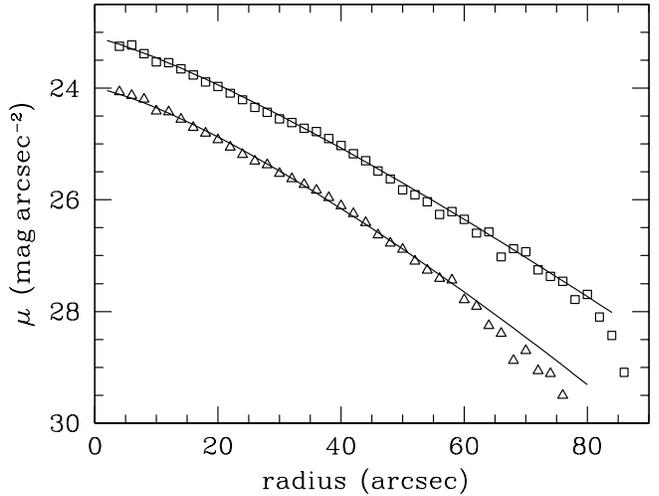}}
\caption[F9.ps]{
Radial $V$ (triangles) and $I$-band (boxes) surface brightness profiles of ESO 540-032.
The solid lines are the best-fit S\'ersic profiles using the model
parameters in Table \ref{props}.
\label{sbprofiles}}
\end{figure}

We cleaned the $VI$ galaxy images from foreground stars and background galaxies using 
procedures written within the IRAF package. Contaminating objects in the vicinity of 
the galaxy were replaced by patches of plain sky. If an object affected the galaxy 
light, the area was replaced instead by a nearby uncontaminated patch from the 
same surface brightness range. The coordinates of the center of the 
luminosity-weighted light distribution was adopted as the galaxy center. Using the 
IRAF command {\it ellipse}, the mean ellipticity of the galaxy was 
measured to be $e=0.20$. We performed simulated aperture photometry on the clean 
images to produce growth curves and $VI$ surface brightness profiles 
(Fig.~\ref{sbprofiles}) as a function of the semi-major axis. The growth curve that 
converges best to a plateau at large distances from the galaxy determined the total 
magnitude $m_{\rm T}$ 
and the magnitude uncertainty was estimated by varying the sky brightness. At half of 
the asymptotic intensity we read off the half-light (``effective'') radius, $r_{\rm eff}$, 
and calculated the mean surface brightness within $r_{\rm eff}$: the ``effective surface 
brightness'' $\langle \mu\rangle_{\rm eff}$. All photometric and structure 
parameters for ESO 540-032 are listed in Table~\ref{props}. 

The observed surface brightness profiles clearly deviate from an exponential 
light profile (a straight line) showing a continuous flattening towards the center. 
Thus, generalized exponential functions (S\'ersic 1968): $I(r)=I_0\exp[-(r/r_0)^n]$, 
with a free shape parameter $n$, were fitted to the data. The inner and outer cut-offs 
for the fits were chosen to be at 3 and 80\,arcsec, respectively. The best-fitting 
S\'ersic profiles are shown as solid lines in Fig.~\ref{sbprofiles} and the 
corresponding model parameters, i.e. the scale length $r_0$, the central surface 
brightness $\mu_0$, and the shape parameter $n$ listed in Table~\ref{props}. 
The overall errors quoted include the profile fitting error and the error due to 
the uncertainty in the sky determination. 

The shape parameters in $V$ and $I$ are consistent with those measured in other 
passbands ($n_B=1.21 \pm 0.07$ and $n_R=1.36 \pm 0.03$, Jerjen et al.~2000b). 
The scale length $r_0$ increases slightly to redder passbands from 
$r_{B,0}$=19.9\, arcsec to $r_{I,0}$=24.6\,arcsec which can be explained
by the high concentration of blue stars in the galaxy center (Sect.~3).  

From the TRGB distance, we derive a dereddened absolute $V$ magnitude
$M_V=-12.1$\,mag and a $V$ central surface brightness of $24$\,mag arcsec$^{-2}$.
If plotted versus the galaxy's mean metallicity $-1.7\pm0.3$\,dex, these data 
points further support the two relations between metallicity and $M_V$ and 
metallicity and central surface brightness defined by the dwarf elliptical galaxies 
in the Local Group (Caldwell et al.~1998; Caldwell 1999).

\begin{table}
\caption[]{Properties of ESO 540-032}
\label{props}
\begin{tabular}{lc}
\hline \hline
  Type                    &             dE/Im (Sph/dIrr)  \\
  R.A.  (J2000.0)              &    $00^h50^m24.6^s$      \\
  Decl. (J2000.0)              &    $-19^\circ 54' 23''$  \\
  Galactic $l$   (deg)       &     $121.00962$            \\
  Galactic $b$   (deg)        &    $-82.77422$            \\
  SuperGalactic $l$ (deg)     &   $276.95088$             \\
  SuperGalactic $b$ (deg)     &   $-4.24336$              \\
  $e$                     &              $0.20$           \\
  $V_T$       (mag)       &              $15.58 \pm 0.12$ \\
  $I_T$       (mag)       &              $14.55 \pm 0.17$ \\
  $E(B-V)$    (mag)       &              0.020            \\
  Extinction: $A_V, A_I$ (mag)  & 0.068, 0.040            \\
  $r_{V, \rm eff}$                  (arcsec)            &   $27.3\pm 1.8$    \\
  $\langle \mu \rangle_{V, \rm eff}$  (mag arcsec$^{-2}$) & $24.8\pm 0.08$ \\
  $r_{I, \rm eff}$                  (arcsec)            &   $29.4\pm 2.2$    \\
  $\langle \mu \rangle_{I, \rm eff}$  (mag arcsec$^{-2}$) &   $23.9\pm 0.09$ \\
  $\mu_{V,0}$                   (mag arcsec$^{-2}$) &   $24.0\pm 0.2$   \\
  $r_{V,0}$                     (arcsec)            &   $23.6\pm 2.7$   \\
  $n_V$                                         &   $1.30\pm 0.12$  \\
  $\mu_{I,0}$                   (mag arcsec$^{-2}$) &   $23.1\pm 0.1$   \\
  $r_{I,0}$                     (arcsec)            &   $24.6\pm 1.6$   \\
  $n_I$                                         &   $1.23\pm 0.06$  \\
  $I_{\rm TRGB}$  (mag)          &           $23.48 \pm 0.09$     \\
  $(m-M)_0$  (mag)          &           $27.64 \pm 0.14$          \\
  Distance (Mpc)            &           $3.4 \pm 0.2$          \\
  $\langle$[Fe/H]$\rangle$ (dex)            &     $-1.7\pm0.3$            \\
\hline \hline
\end{tabular}
\end{table}

\section{Globular Cluster Candidate}
\begin{figure}
\centering
\resizebox{\hsize}{!}{
\includegraphics{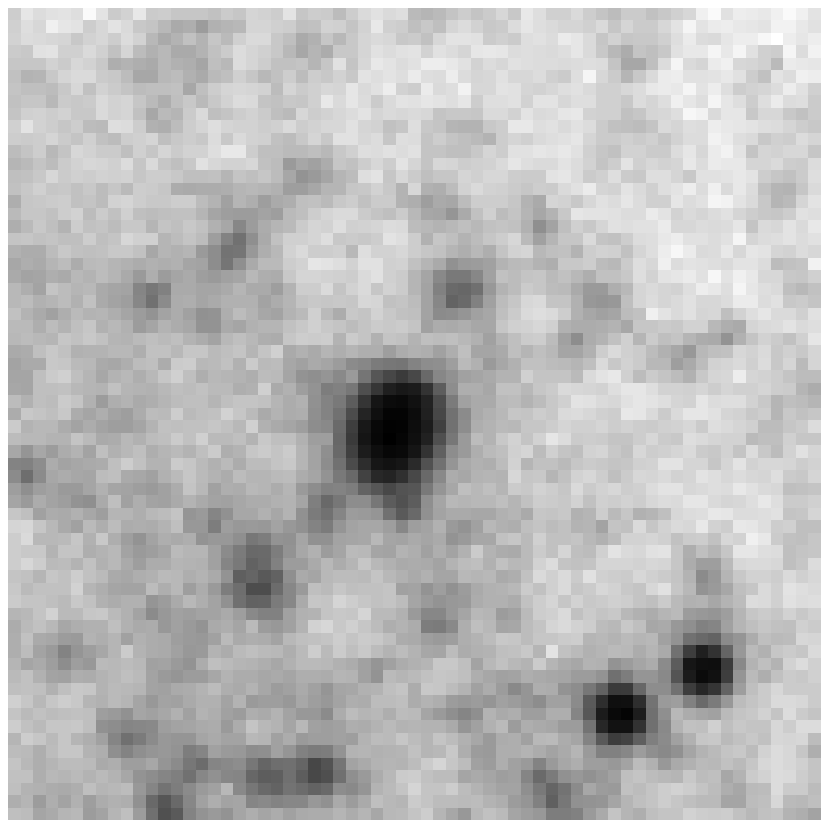}
\includegraphics{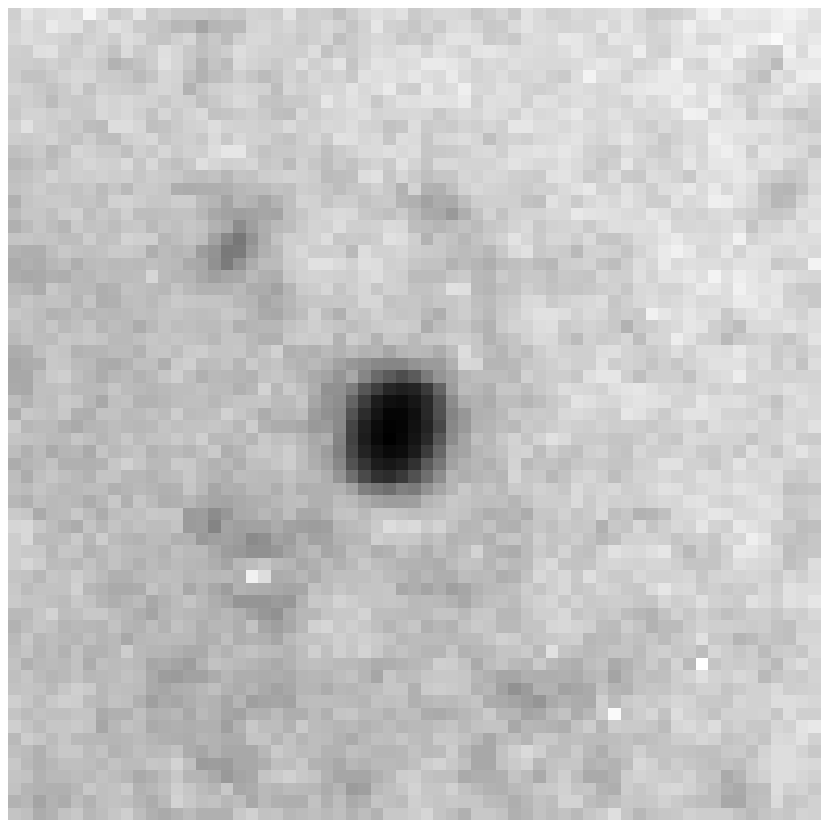}
}
\caption{
The $V$ image of the globular cluster candidate before (left) and after (right) 
the subtraction of neighbouring stars. The cleaned image was produced by the 
allstar task in IRAF using the stellar PSF. The size of the images is 
$12\farcs8 \times 12\farcs8$. North is up, east to the left. 
\label{gccand}}
\end{figure}

While reducing the photometric data, we noticed a non-stellar object  
at the north eastern part of the galaxy ($00^h 50^m 24.1^s$, $-19^\circ 54'
15''$, J2000), $11\farcs2$ from the center (see Fig.\ref{gccand}). It has 
a profile that is more globular cluster (GC) like in the sense that the light 
distribution has broad wings with a dispersion 
$\sigma=1.18$\,arcsec. The mean FWHM of a star on the CCD frame is 
$\sim0.6$\,arcsec. If this object is indeed a GC,
we get a rough estimate of its core radius $r_c$ with the form 
$2r_c=\sqrt{\sigma_{GC}^2- \sigma_{star}^2} = 1.02$\,arcsec. 
Employing our TRGB distance this translates into $r_c=8.4$\,pc. 
For comparison, the core radius of $\omega$ Cen, the most luminous
GC in the Milky Way has only a core radius of 3.8\,pc (Harris 1996). However,
there are several Galactic GCs that have core radii even
larger than the one we estimate from our images (e.g.
NGC 5053, Pal5, Pal14, Pal145, Pyxis; see Harris' web 
catalogue\footnote{http://physun.physics.mcmaster.ca/Globular.html}).
$(V-I)_0$ colour as well as $V$ and $I$ magnitudes of the extended object 
[$(V-I)_0=0.9$, $V_0=-7.5$ and $I_0=-8.4$ mag] are within the
range observed for the MW globular clusters (Harris 1996).

Alternatively, the unknown object could well be an \ion{H}{ii} region like 
the one identified in ESO 294-010 (JFB98) or a galaxy located in the
background of ESO 540-032 similar to the case of BK5N, a dwarf in the M81
group, where an apparent nucleus was in fact a distant spiral galaxy
(Caldwell et al.~1998).

\section{Summary and Conclusions}
Based on CCD images obtained at the ESO Very Large Telescope UT1+FORS1, 
we presented the $(I,V-I)$ colour-magnitude diagram and the $I$-band luminosity 
function for the nearby dwarf galaxy ESO 540-032. The CMD shows traces  
of two distinct stellar components, a predominant population of 
metal-poor ([Fe/H]$\approx -1.7$\,dex) stars of ages similar to those of 
globular clusters and a small fraction of young ($150-500$\,Myr), more 
metal-rich ([Fe/H]$\approx -1.3$\,dex) stars. In addition, the CMD reveals a 
considerable number of stars brighter than the RGB tip. They could be AGB 
stars and represent an intermediate-age population. In this case, ESO 540-032 
would have had a continuous star formation activity like the Fornax dwarf. 
Alternatively, these stars are foreground stars, a scenario which is supported 
by the CMDs of comparison fields and thus more likely. The young stars in 
ESO 540-032 would then be the result of a single event comparable to the situation 
observed in the two Local Group dwarfs Phoenix and LSG3.  

The tip of the red giant branch magnitude $I_0$ = $23.48\pm$0.09\,mag derived 
from the $I$-band luminosity function of ESO 540-032 corresponds to a true 
distance modulus of $27.64 \pm 0.14$\,mag, or $D=3.4 \pm 0.2$\,Mpc. This 
new distance for ESO 540-032 seems to be more reliable than the previously published 
value of $D=2.2 \pm 0.14$\,Mpc. The latter result was based on a successful measurement of 
the surface brightness fluctuation signal in the galaxy but relied on incorrect assumptions 
about the stellar mixture. From our experiments with Worthey's stellar population 
synthesis models, we draw the conclusion that there is no advantage for the SBF method 
over the TRGB method to measure distances of intermediate-type (dE/Im) dwarf galaxies. 
Applications of the SBF technique should be concentrated on genuine dwarf elliptical 
and dwarf S0 galaxies. 

In space, ESO 540-032 is located in the field behind the spiral galaxy NGC 247, a region 
which is part of the elongated nearby cloud of galaxies in Sculptor. Its $VI$ surface 
brightness profiles have bee derived and fitted by S\'ersic models. The mean 
metallicity, the total $V$ luminosity and the central surface brightness in 
$V$ follow closely the relations exhibit by the early-type dwarf galaxies in the 
Local Group. In ESO 540-032, we are likely to witness the final stage of the transition 
from a dwarf irregular to a dwarf elliptical galaxy. Finally, we discussed the 
properties of an extended object in the central area of ESO 540-032. Colour, physical 
size, and absolute magnitudes do not rule out that it is a large globular cluster what 
we see. 

The first priority of future work will be to measure colour-magnitude diagrams 
for the remaining four dwarfs NGC 59, Scl-dE1, ESO 294-010, and ESO 540-030 
of the JFB98 sample in the Sculptor group region. They will allow to determine 
metallicities and metallicity spreads and to search for evidence of recent star 
formation activities from blue-loop stars. The tip of the red giant branch distances 
will provide a check of the existing SBF distances. If the data are combined with 
radial velocities, these dwarfs can be used as additional test particles for 
dynamical models of the Sculptor--Local Group cloud which will help to refine 
our knowledge of this complex structure in the Supergalactic plane. 

\begin{acknowledgements}
We are grateful to the anonymous ESO service observer at the VLT UT1 (Antu) for 
providing excellent quality images for this study. It is a pleasure to thank 
G.~Da Costa and B.~Binggeli for useful discussions and the referee G.P.~Bertelli 
for interesting comments. HJ acknowledges the financial support from the {\em 
Swiss National Science Foundation}. 
\end{acknowledgements}

\end{document}